\documentclass[10pt,twocolumn]{research-article-2col}

\templatetype{research-article-2col} 

\usepackage{rotating}  
\usepackage{booktabs}
\usepackage{tablefootnote}
\usepackage{xcolor}
\usepackage{sidecap}

\newcommand{\na}{\textsuperscript{23}Na }    
\newcommand{\nax}{\textsuperscript{23}Na}    

\newcommand{\hhna}{\textsuperscript{1}H/\textsuperscript{23}Na }

\newcommand{\ttwo}{T\textsubscript{2} }
\newcommand{\ttwox}{T\textsubscript{2}}

\newcommand{\ttwoshort}{T\textsubscript{2,short} }

\newcommand{\ttwolong}{T\textsubscript{2,long} }

\title{Analysis of blurring due to short T\textsubscript{2} decay at different resolutions in \textsuperscript{23}Na MRI}

\author[1]{Olga Dergachyova}
\author[1,2]{Zidan Yu} 
\author[1,3,4]{Shota Hodono} 
\author[1,3]{Martijn Cloos} 
\author[1,*]{Guillaume Madelin} 

\affil[1]{Center for Biomedical Imaging, Department of Radiology, New York University Grossman School of Medicine, New York, NY, USA}
\affil[2]{Department of Medicine, John A. Burns School of Medicine, University of Hawaii, Honolulu, HI, USA}
\affil[3]{Centre for Advanced Imaging, The University of Queensland, Brisbane, QLD, Australia}
\affil[3]{Donders Centre for Cognitive Neuroimaging, Donders Institute for Brain, Cognition and Behaviour, Radboud University, Nijmegen, The Netherlands}
\affil[*]{Corresponding author: guillaume.madelin@nyulangone.org}

\begin{abstract}

The nuclear magnetic resonance signal from sodium (\nax) nuclei demonstrates a fast bi-exponential \ttwo decay in biological tissues (\ttwoshort = 0.5-5 ms and \ttwolong = 10-30 ms). Hence, blurring observed in sodium images acquired with center-out sequences is generally assumed to be dominated by signal attenuation at higher k-space frequencies. Most of the studies in the field primarily focus on the impact of readout duration on blurring but neglect the impact of resolution. In this paper, we examine the blurring effect of short \ttwo on images at different resolutions. A series of simulations, as well as phantom and in vivo scans were performed at varying resolutions and readout durations in order to evaluate progressive changes in image quality. We demonstrate that, given a fixed readout duration, \ttwo  decay produces distinct blurring effects at different resolutions. Therefore, in addition to voxel size-dependent partial volume effects, the choice of resolution adds additional \ttwox-dependent blurring.

\end{abstract}

\dates{\textit{\today}}

\begin{document}

\maketitle


\section{Introduction}

Sodium (\nax) magnetic resonance imaging (MRI) can be used to non-invasively study metabolic processes and has potential for assessing early stages of neurodegenerative diseases, as well as muscular channelopathies, cancer malignancy and therapy \cite{madelin2013}. Because of the relatively small concentration of sodium ions (Na\textsuperscript{+}) in human body (of the order of tens to hundreds of mmol/L, or mM), even high-field sodium MRI results in images with low resolution and low signal-to-noise ratio (SNR). In addition, in many biological tissues, such as gray matter (GM) and white matter (WM) in brain, the \na MR signals undergo fast bi-exponential transverse (\ttwox) relaxation (\ttwoshort = 0.5-5 ms and \ttwolong = 10-30 ms) \cite{madelin2013, boada1994}. 

To capture as much signal as possible, ultra-short echo time (UTE) pulse sequences with center-out trajectories in k-space are often used in sodium MRI. When using a center-out trajectory, the center of k-space, corresponding to an overall image contrast is filled first, and its outer parts, corresponding to high spatial frequencies and image contours, are acquired last. These sequences usually follow non-Cartesian radial or spiral trajectories that must be reconstructed using a non-uniform fast Fourier transform (NUFFT) \cite{fessler2003}. Examples of sodium center-out sequences include radial projections (RP) \cite{jereric2004}, twisted projection imaging (TPI) \cite{boada1997a}, 3D cones \cite{staroswiecki2010}, spiral trajectories \cite{gai2015}, Fermat looped orthogonally encoded trajectories (FLORET) \cite{pipe2011}, and twisting radial lines (TWIRL) \cite{konstandin2015}. 

Higher k-space frequencies in center-out sequences suffer from signal attenuation \cite{robson2003} because of the fast decay of short \ttwo components. Such \ttwox-induced signal attenuation is well studied in standard proton (\textsuperscript{1}H) MRI. In Cartesian trajectories, it is known to cause blurring along the phase encoding direction when using sequences with long echo trains \cite{qin2012} such as FSE \cite{constable1992}, HASTE \cite{zhang2009}, MPRAGE \cite{deichmann2000}, or EPI \cite{oshio1989}. Besides Cartesian trajectories, blurring due to signal attenuation in \textsuperscript{1}H imaging was also reported for center-out spiral EPI \cite{ahn1986} and 3D radial trajectory used to image short \ttwo components \cite{rahmer2006}.

Because of the fast \ttwo relaxation of \nax, it is commonly believed that blurring in sodium images is also a result of signal attenuation and can be reduced using shorter readouts \cite{nielles2007, boada1997b, wang2009, wheaton2004, bottomley2016, stobbe2018}. However, short readouts lead to low SNR. Hence, an optimal readout duration should be chosen to reach a balance between SNR and blurring. Some density-adapted sampling schemes were proposed for TPI \cite{boada1997b} and RP \cite{nagel2009} to reduce signal attenuation by performing more efficient and uniform sampling. Nevertheless, mitigation of \ttwox-induced signal attenuation and blurring in \na imaging is still an area of ongoing research.

Several studies evaluated the impact of different types of trajectories used for \na MRI based on resulting blurring, SNR and spatial resolution \cite{konstandin2015, rahmer2006, boada1997b, nagel2009, romanzetti2014}. Yet, they mostly explored the influence of sampling schemes and variable readouts. Thus, currently, readout duration is the parameter of choice when it comes to optimizing sodium image quality and finding a good balance between SNR and blurring. 

\begin{figure*}[t]
    \centering
    \includegraphics[width=1\textwidth]{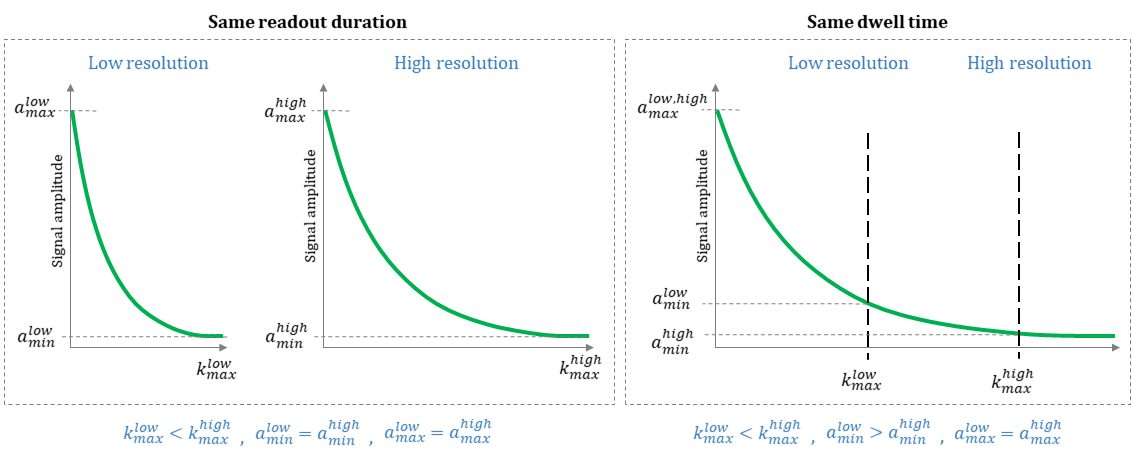}
    \caption{Signal T\textsubscript{2} decay (magnitude amplitude from $a_{max}$ to $a_{low}$) in k-space in case of a fixed readout duration (left panel) and a fixed dwell time (right panel) for low and high resolutions, along the same center-out radial or spiral projection (i.e., with the same gradient amplitude). As a reminder: $k_r(t) = \int_{0}^{t} \gamma G_r(\tau) d\tau$, with $k_r(t)$ the position of the trajectory in k-space related to spatial direction $r$ at time point $t$, $\gamma$ the gyromagnetic ratio of the nucleus of interest (in our case, \textsuperscript{23}Na has $\gamma = 7.0808493\times10^7$ rad T\textsuperscript{-1} s\textsuperscript{-1}), and $G_r(t)$ the linear magnetic gradient in direction $r$ at time point $t$. For example, in the case of a simple radial center-out projection, readout duration $T=N\cdot dw$, with $N$ the total number of data points acquired along the projection with dwell time $dw$, such that $k(t) = \gamma G t$, with $t=n\cdot dw$ and $n=[1,N]$, with $G$ the amplitude of a constant gradient, and $k_{max} = \gamma G T$.}
    \label{fig:kspace}
    \vspace{-12pt}
\end{figure*}

Based on our own observations that led to the idea of this paper, we suspected that the blurring effect from fast \ttwo decay of the signal is different at different resolutions and does not depend on readout duration alone but also on k\textsubscript{max} (inverse of spatial resolution). Even though, in reality, a \na nucleus undergoes the same relaxation and emits the same signal regardless of the readout duration and resolution, the information about its signal decay is stored in k-space differently. Because k\textsubscript{max} depends on the chosen resolution of the image, the decay curve of a signal acquired with a fixed readout duration will have the same maximum and minimum, but will be stretched out or squeezed down depending on the extent of the k-space, as shown in \textbf{Figure \ref{fig:kspace}}. We hypothesized that this would result in different \ttwo blurring effects for different resolutions. The shape of the signal's curve is preserved in the case of a fixed dwell time (i.e., time required to acquire one k-space sample), but variable readout duration, which results in the same \ttwo blurring effect for all resolutions. However, selecting dwell time as one of the key acquisition parameters is an unpopular choice in practice.

To the best of our knowledge, no extensive analysis has been conducted to explore the influence of different resolutions on blurring compared to readout duration. Therefore, the purpose of this paper is to explore how the choice of resolution impacts blurring in comparison to readout duration in the context of \na brain MRI using a 2D radial center-out trajectory. The paper presents both quantitative and qualitative comparative evaluation of blurring related to these parameters through a series of simulations and acquisitions performed on a home-made "brain-like" phantom and in brain in vivo.

\vfill

\section{Material and Methods}

\subsection{Simulations}

To quantitatively measure blurring, we proposed several experiments conducted in simulations. In these experiments, we simulated the \ttwo decay function and 2D radial central-out trajectory point spread function (PSF), and their combination, when using different acquisition parameters, namely resolution, readout duration and dwell time. Relative amplitude and full width at half maximum (FWHM) of the PSF were used to compare these blurring functions. 

\subsubsection{Mono-exponential decay blurring function \textit{f}} \label{sec:methods_mono}

We decided to start out experiments by exploring the blurring induced by mono-exponential decay as a function of \ttwo and at different resolutions and readout durations. In this way, a change can be observed for a range of hypothetical tissues and scan settings. In sodium MRI many tissues produce a bi-exponential decay due to the magnetic properies of \na spin $\frac{3}{2}$, but it can be thought mathematically of as a superposition of two mono-exponential fits. Thus evaluating a mono-exponential decay will help understand existing dependencies more easily. In \cite{rahmer2006}, Rahmer et al. proposed an analytical formula for \ttwo blurring function $f(r)$ corresponding to a 2D radial center-out trajectory. We used it to simulate a blurring kernel that, by convolving it with a true image of an object in image space, would produce the blurring induced by \ttwo decay. In this formula, shown in \textbf{Eq. (\ref{eq:blurring})}, variable \textit{r} denotes radius in spherical coordinates in image space and \textit{T\textsubscript{read}} corresponds to readout duration:
\begin{equation}
    f(r) = \cfrac{2 \left( \cfrac{T_2}{T_{read}} \right)^2}
    {\left( 1 + \left( rk_{max}\cfrac{T_2}{T_{read}} \right)^2 \right)^{\frac{3}{2}}}
    \label{eq:blurring}
\end{equation}


The formula was computed for the following \ttwo values: 2, 4, 10, 20, 40, 60 ms, and the readout duration was varied from 5 to 25 ms with a step of 5 ms. It was also computed for ten resolutions with square pixels of size 0.1, 0.2, 0.3, 0.4, 0.5, 1, 2, 3, 4, 5 mm. The FWHM of $f(r)$ was computed for all the combinations (in mm), and its progression is shown for a selection of parameters to support the analysis.

\subsubsection{Bi-exponential decay blurring function \textit{f}} \label{sec:methods_bi}

To approach more realistic scan settings, such as bi-exponential decay and lower resolutions (above 1$\times$1 mm\textsuperscript{2}), we performed the following set of simulations. 

First, we adapted the formula from \cite{rahmer2006} to bi-exponential decay with \ttwoshort = 3 ms and \ttwolong = 22 ms with 60\% and 40\% signal proportion, respectively (average values and proportions from the literature \cite{madelin2013,boada1994,stobbe2018}).

Secondly, we evaluated the blurring of $f(r)$ from bi-exponential decay in two cases: (1) a set of readout durations, and (2) a set of dwell times. Usually, the readout duration is the parameter of choice to optimize image quality in terms of SNR and blurring. Because of the phenomenon shown in \textbf{Figure \ref{fig:kspace}}, it is hard to distinguish between individual contributions of \ttwo and resolution to image blurring. That is why selecting dwell time instead of readout duration can help observe the resolution-invariant blurring effect of short \ttwo decay.

In this simulation, four in-plane resolutions were tested: 1$\times$1 mm\textsuperscript{2}, 2$\times$2 mm\textsuperscript{2}, 3$\times$3 mm\textsuperscript{2}, and 4$\times$4 mm\textsuperscript{2} pixels. As in section \ref{sec:methods_mono}, the readout duration varied from 5 to 25 ms with a step of 5 ms. Dwell time was varied from 200 to 1000 $\mu$s with a 200 $\mu$s step. The comparison of signal's relative amplitude and blurring function's FWHM was made for both cases.

\subsubsection{Trajectory \textit{PSF}}

Sampling signal using a non-Cartesian trajectory requires a NUFFT reconstruction \cite{fessler2003}. Discrete sampling, possible angular undersampling, and gridding process used in NUFFT lead to a modified PSF. To obtain an image space representation (i.e., blurring kernel) of trajectory's PSF, we first filled a k-space of an appropriate size (depending on the resolution) with a constant value of `1' along the radial projections, and then took its inverse NUFFT. The operation was performed for the same four resolutions as in section \ref{sec:methods_bi}.

\subsubsection{Combined blurring function \textit{B}}

We combined the bi-exponential decay function and the trajectory's PSF to obtain a final blurring function \textit{B}. This was done by convolving the PSF kernel of the trajectory by the computed bi-exponential decay function $f(r)$ in image space (\textit{B = PSF $*$ f}), which corresponds to their multiplication in k-space. The operation was performed for all the parameters stated in section \ref{sec:methods_mono}.

\subsection{Phantom experiments}

To empirically validate theoretical results of the simulation and provide a visual comparison, we performed two experiments on a home-made sodium phantom. 

The first experiment, described in section \ref{sec:methods_readout}, placed the problem in a conventional context where readout duration is considered to be the main parameter for (de)blurring optimization. In this experiment, different combinations of resolutions and readout durations were used to acquire images and display relative influence of both parameters. 

The second experiment, described in section \ref{sec:methods_dt}, was designed to show the pure impact of resolution on blurring: for all resolutions, images were acquired with a same very short dwell time that limits the effect of fast \ttwo decay related signal attenuation. All the acquisitions were performed at 7 T (MAGNETOM scanner, Siemens, Erlangen, Germany) using a 16-channel Tx/Rx \hhna coil developed in-house \cite{wang2021radially}. 

\subsubsection{Phantom preparation and relaxation times} \label{sec:methods_phantom}

The phantom was designed to have Na\textsuperscript{+} concentrations and relaxation properties similar to human brain. It contained four compartments that, from the outside to the inside, were designed to roughly imitate cerebrospinal fluid (CSF) around the brain, gray matter (GM), white matter (WM), and ventricles also containing CSF. Both CSF compartments were filled with a 140 mM NaCl solution, whereas the GM and WM compartments were filled with 2\% agar gels with 40 mM and 35 mM NaCl concentrations, respectively. The frame for the phantom was 3D printed (Form 2, Formlabs, Somerville, MA, USA) with 1 mm-thick walls separating the compartments. 

\begin{table*}[t]
    \caption{Number of spokes and dwell times (rounded to integers for display only) at different resolutions and readout durations for the first experiment on phantom.}
    \centering
    \small
    \vspace{-6pt}
    \begin{tabular}{m{46mm}m{20mm}m{20mm}m{20mm}m{20mm}m{20mm}}
        \hline
        \textbf{Readout duration} & \textbf{5 ms} & \textbf{10 ms} & \textbf{15 ms} & \textbf{20 ms} & \textbf{25 ms} \\
        \hline
        \textit{Resolution} & \multicolumn{5}{c}{\textbf{Number of spokes}} \\
        \hline
        1$\times$1 mm\textsuperscript{2} & 24,000 & 16,970 & 13,855 & 12,000 & 10,735 \\
        2$\times$2 mm\textsuperscript{2} & 4,245 & 3,000 & 2,450 & 2,120 & 1,895 \\
        3$\times$3 mm\textsuperscript{2} & 1,540 & 1,090 & 890 & 770 & 690 \\
        4$\times$4 mm\textsuperscript{2} & 750 & 530 & 430 & 375 & 335 \\
        \hline
        \textit{Resolution} & \multicolumn{5}{c}{\textbf{Dwell time}} \\
        \hline
        1$\times$1 mm\textsuperscript{2} & 52 $\mu$s & 104 $\mu$s & 156 $\mu$s & 208 $\mu$s & 260 $\mu$s \\
        2$\times$2 mm\textsuperscript{2} & 104 $\mu$s & 208 $\mu$s & 313 $\mu$s & 417 $\mu$s & 521 $\mu$s \\
        3$\times$3 mm\textsuperscript{2} & 156 $\mu$s & 313 $\mu$s & 469 $\mu$s & 625 $\mu$s & 781 $\mu$s \\
        4$\times$4 mm\textsuperscript{2} & 208 $\mu$s & 417 $\mu$s & 625 $\mu$s & 833 $\mu$s & 104 $\mu$s \\
        \hline
    \end{tabular}
    \vspace{-6pt}
    \label{tab:params}
\end{table*}

To confirm that the phantom had appropriate relaxation properties, we performed 16 scans using a FLORET sequence \cite{pipe2011,madelin2014} to measure T$^*_2$ values of the phantom compartments. We used 3 hubs with 45$^{\circ}$ angle, 100 interleaves, 60$^{\circ}$ flip angle, 256$\times$256$\times$256 mm\textsuperscript{3} field of view (FOV), 4$\times$4$\times$4 mm\textsuperscript{3} pixel size, 6 averages, 150 ms TR, and 16 different exponentially spaced TE (0.1, 0.2, 0.5, 0.8, 1.2, 1.6, 2, 3, 5, 8, 12, 18, 25, 35, 50, and 65 ms) with a total acquisition time of 144 s for each scan. For each TE, we extracted three regions of interest (one per compartment, where the two CSF compartments were considered as one) from the axial plane and computed their average signal magnitudes normalized by the maximum image magnitude. Then, a bi-exponential T$^*_2$ decay curve from \textbf{Eq. (\ref{eq:decay})} was fitted to each compartment. The fit was performed in Matlab using non-linear regression (\emph{nlinfit}). In \textbf{Eq. \ref{eq:decay}}, $M_{xy}$ and $t$ were known parameters corresponding to average signal magnitudes and TE respectively. The parameters to fit were $T^*_{2,short}$ and $T^*_{2,long}$ corresponding to short and long transverse relaxation times, $a_{short}$ and $a_{long}$ corresponding to their respective proportions, with $a_{short} + a_{long} = 1$, and additive noise $N$:
\begin{equation}
    M_{xy} = M_0 \left( a_{short}e^{-t/T^*_{2,short}} + a_{long}e^{-t/T^*_{2,long}} \right) + N
    \label{eq:decay}
\end{equation}

\subsubsection{First experiment: Variable readout duration} \label{sec:methods_readout}

This experiment corresponded to simulations from section \ref{sec:methods_bi}, where both resolution and readout durations were varied to observe their mutual impact on phantom blurring. Here, we acquired sets of images for four in-plane resolutions (1$\times$1 mm\textsuperscript{2}, 2$\times$2 mm\textsuperscript{2}, 3$\times$3 mm\textsuperscript{2}, and 4$\times$4 mm\textsuperscript{2} pixels), and five readout durations (5, 10, 15, 20, and 25 ms). 

All the images were acquired using a 2D center-out radial GRE sequence with the following parameters: 10 mm-thick axial slices, 192$\times$192 mm\textsuperscript{2} FOV, 65 ms TR, 0.7 ms TE, and 60$^{\circ}$ flip angle. The number of samples per readout were 96, 48, 32, or 24. The readout duration was held constant, i.e., the dwell time was increased with decreasing resolution. Given that shorter dwell times and higher resolutions decrease the SNR, we increased the number of spokes (which is similar to averaging in this case) to insure a similar SNR was maintained in all images. See details in \textbf{Table \ref{tab:params}}. 

\subsubsection{Second experiment: Short dwell time} \label{sec:methods_dt}

Experimentally, resolution-related blurring is difficult to separate from blurring due to \ttwo decay. However, it is possible to emphasize the effect of resolution itself and minimize the effect of fast \ttwo decay. For this, we acquired four sets of images (one per resolution) at one fixed very short dwell time of 8.5 $\mu$s. The chosen resolutions are the same as in the first experiment. 

\begin{figure*}[t!]
    \centering
    \includegraphics[width=1\textwidth]{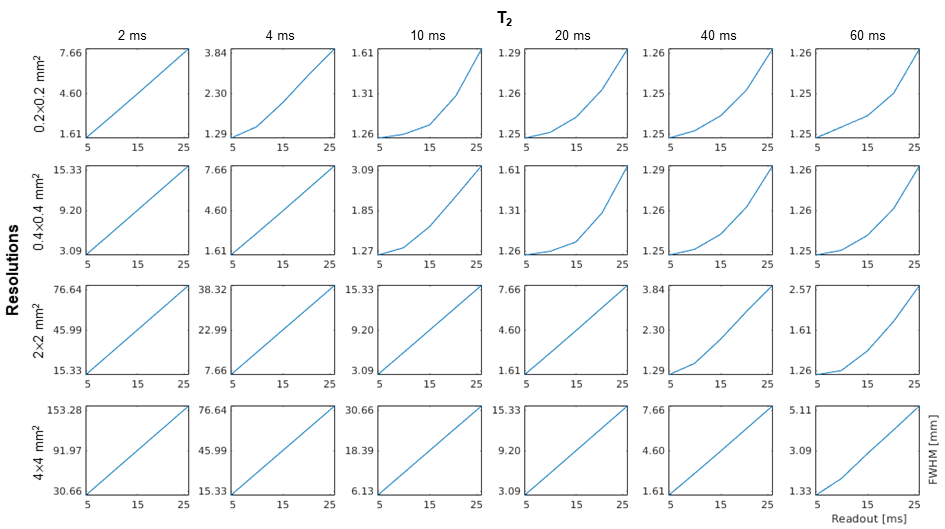}
    \caption{Representative examples of progression of the full width at half maximum (FWHM, in mm) of the mono-exponential decay blurring function $f(r)$ for different readout durations, resolutions (or k\textsubscript{max}), and \ttwo values. See Section \ref{sec:methods_mono}.}
    \label{fig:mono}
    \vspace{-12pt}
\end{figure*}

The same radial GRE sequence was used with the following common parameters: 10 mm slices, 192$\times$192 mm\textsuperscript{2} FOV, 15 ms TR, 0.8 ms TE, 45$^{\circ}$ flip angle, and 5 averages. As before, the number of spokes per acquisition was varied to maintain constant SNR. From the highest resolution to the lowest, the number of samples was set to 96, 48, 32, and 24, which corresponded to 816, 408, 272, and 204 $\mu$s readout durations. The number of spokes was set to 24,000, 6,000, 2,667, and 1,500 respectively to accommodate for change in voxel volume and maintain similar SNR. 

\subsection{In vivo brain experiment}

Exploring the same range of acquisition parameters in vivo would require impractically long scan times. Therefore, we decided to limit ourselves to only one readout duration and four resolutions, using the same sequence as for the phantom scans. A readout duration of 15 ms was chosen as a good balance between potential blurring and total scan time (i.e., number of spokes) required to obtain a sufficient SNR. Other acquisition parameters were identical to the ones provided in section \ref{sec:methods_readout}, except that this time 2 averages were acquired for each resolution. The brain of a healthy female volunteer was scanned for a total of 30 min, 5 min 18 s, 1 m 56 s, and 56 s for 1$\times$1 mm\textsuperscript{2}, 2$\times$2 mm\textsuperscript{2}, 3$\times$3 mm\textsuperscript{2}, and 4$\times$4 mm\textsuperscript{2} resolutions, respectively. The study was approved by our institutional review board (IRB), and written informed consent was obtained prior to examination. All the in vivo acquisitions were also performed at 7 T using a 16-channel Tx/Rx \hhna coil \cite{wang2021radially}. 


\subsection{Reconstruction and B\textsubscript{0} correction}

The images from phantom and in vivo experiments were reconstructed in MATLAB using the Fessler's NUFFT toolbox \cite{fessler2003} for non-Cartesian data sampling. No data filtering in k-space was performed prior to reconstruction to avoid any added blurring. 

Center-out radial acquisitions with long readout durations are susceptible to additional blurring induced by local B\textsubscript{0} inhomogeneities. To mitigate these effects, B\textsubscript{0} maps were also acquired for all resolutions to correct the phantom images for B\textsubscript{0} inhomogeneity using the method described described in \cite{noll1992}. Using this method, for each resolution, the off-resonance values from its B\textsubscript{0} map were divided into 50 equal intervals. The intervals were transformed into a list of values, where each list entry represented the smallest value in the corresponding interval. For every value in the list, an image was reconstructed from the signal corrected with the corresponding resonance offset. To obtain the final image, every pixel was taken from a reconstructed image corrected with the closest off-resonance value. 

\begin{figure*}[t]
    \centering
    \includegraphics[width=1.0\textwidth]{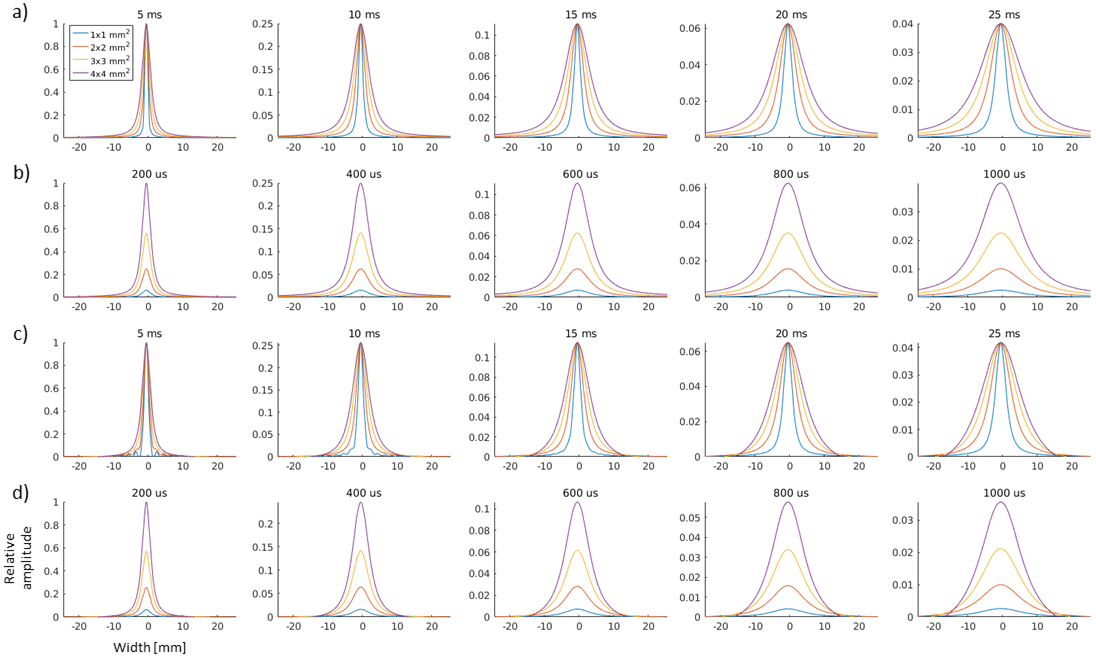}
    \caption{Blurring functions: decay-related blurring functions $f$ for different readout durations (row \textit{a}), decay-related blurring functions $f$ for different dwell times (row \textit{b}), combined blurring functions \textit{B = PSF $*$ f} (convolution of decay-related blurring functions and trajectory PSFs) for different readout durations (row \textit{c}), and combined blurring functions $B$ for different dwell times (row \textit{d}). Relative amplitudes were normalized by the highest amplitude in a row.}
    \label{fig:bi_total}
    \vspace{-12pt}
\end{figure*}

\vfill

\section{Results}

\subsection{Simulations}

\subsubsection{Mono-exponential decay function \textit{f}}

\textbf{Figure \ref{fig:mono}} shows the changes in FWHM of \textit{f} through 5 readout durations (5, 10, 15, 20, and 25 ms), for 5 different \ttwo values (2, 4, 10, 20, 40, and 60 ms), and a representative selection of 4 resolutions (0.2$\times$0.2, 0.4$\times$0.4, 2$\times$2, and 4$\times$4 mm\textsuperscript{2} pixel sizes). 


We first observed that, at any given resolution: (1) for the same readout duration, the FWHM of the blurring function \textit{f} increased when T$_2$ shortened; and (2) when readout durations increased, this FWHM became wider faster when T$_2$ shortened. For example, for 1$\times$1 mm\textsuperscript{2} pixels (not shown) and the shortest \ttwo (2 ms), the FWHM increased from 7.66 to 38.32 mm (by a factor of about 5) throughout different readout durations, while for longest \ttwo (60 ms), the FWHM only increased from 1.26 to 1.45 mm (by a factor of about 1.15). For 2$\times$2 mm\textsuperscript{2} pixels (shown) and the shortest \ttwo (2 ms), the FWHM increased from 15.33 to 76.64 mm (also by a factor of about 5) throughout the different readout durations, while for the longest \ttwo (60 ms), the FWHM only changed from 1.26 to 2.57 mm (by a factor of about 2).

We also observed that the FWHM of the mono-exponential decay blurring function $f$ was always larger for lower resolutions, and that at those lower resolutions, it drastically increased when readout durations increased in case of shorter \ttwox. For example, for the highest resolution (0.1$\times$0.1 mm\textsuperscript{2}) and the longest \ttwo (60 ms), the FWHM was roughly 1.25 mm with a change of less than 0.02\% throughout the readout durations, while for the shortest \ttwo (2 ms) it only doubled, growing from 1.29 to 3.84 mm. On the contrary, for the lowest resolution (5$\times$5 mm\textsuperscript{2}), the FWHM increased from 1.45 to 6.39 mm (by a factor of about 4.4) for the longest \ttwox, and from 38.3 to 191.61 mm (by a factor of about 5) for the shortest \ttwox, when readout durations increased from 5 ms to 25 ms. Therefore, at the longest readout duration (25 ms) and the shortest \ttwo (2 ms), the blurring effect was almost 50 times stronger for 5$\times$5 mm\textsuperscript{2} pixels compared to 0.1$\times$0.1 mm\textsuperscript{2} pixels (FWHM ratio: 191.61/3.84 $\sim$ 49.90), while it was only about 5 times stronger for the longest \ttwo = 60 ms (FWHM ratio: 6.39/1.25 $\sim$ 5.11).


Finally, we observed an exponential-type curvature of the FWHM increase when readout duration increased at very high resolutions and longer \ttwox, whereas this FWHM increase became practically linear at low resolutions.

\vfill

\begin{figure*}[t]
    \centering
    \includegraphics[width=1.0\textwidth]{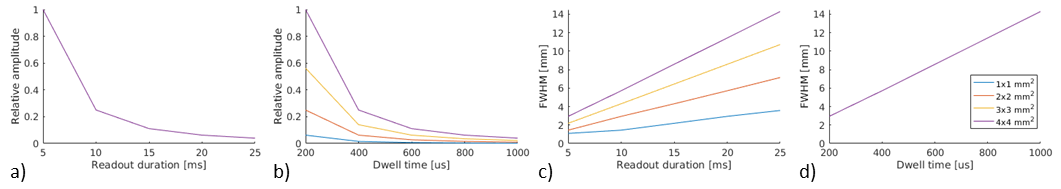}
    \caption{Relative amplitudes and full widths at half maximum (FWHM) for decay-related blurring functions $f$ at variable readout durations (\textit{a} and \textit{c}) and variable dwell times (\textit{b} and \textit{d}).}
    \label{fig:decay_stats}
    \vspace{-12pt}
\end{figure*}

\begin{figure*}[b]
    \centering
    \includegraphics[width=1.0\textwidth]{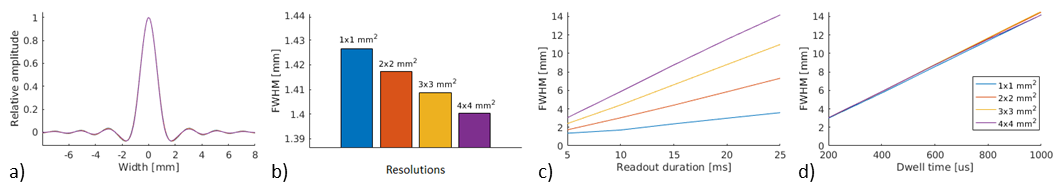}
    \caption{Trajectory point spread functions (PSF) (\textit{a}), their full widths at half maximum (FWHM) (\textit{b}), and FWHM of the combined blurring functions $B$ in cases of variable readout duration (\textit{c}) and variable dwell time (\textit{d}).}
    \label{fig:trajectory_total}
    \vspace{-8pt}
\end{figure*}

\subsubsection{Bi-exponential decay function \textit{f}}

\textbf{Figure \ref{fig:bi_total}.a} presents decay functions in image space for the different readout durations, while \textbf{Figure \ref{fig:bi_total}.b} presents results for different dwell times. For all readout durations, the function was wider for lower resolutions. For example, at 5 ms readout duration, the FWHM for 1$\times$1 mm\textsuperscript{2} pixels was 1.09 mm, while it was 2.94 mm for 4$\times$4 mm\textsuperscript{2} pixels. At 25 ms readout duration, the FWHM was 3.58 mm and 14.28 mm for the same resolutions, respectively. The function amplitude was, however, the same for all resolutions at a given readout duration, and decreased with the increase of readout duration (\textbf{Figure \ref{fig:decay_stats}.a}). \textbf{Figure \ref{fig:decay_stats}.b} presents results for different dwell times. In this case, all functions had different amplitudes that decreased with increased dwell time. Nevertheless, at a given dwell time, the functions had the same width for each resolution, which increased only with dwell time. For each resolution at 200 $\mu$s and 1 ms dwell times, the FWHM was 2.94 mm and 14.28 mm, respectively.

\textbf{Figure \ref{fig:decay_stats}.c} depicts the blurring function's FWHM for different readout durations. Two main observations can be made: (1) low resolution at short readout durations caused a comparable amount of blurring as high resolution at long readout durations, and (2) the blurring from long readout durations was intensified by low resolution. However, \textbf{Figure \ref{fig:decay_stats}.d} confirms our previous hypothesis that blurring effect from \ttwo decay actually depends on dwell time. Simulations with different readout durationss and resolutions while keeping the dwell time fixed resulted in the same FWHM. Nevertheless, the analytical formula used for the simulation does not take into account any other resolution-dependent effects that are not related to transverse relaxation. That is why the FWHM appears to be the same here but can be slightly different in reality due to other resolution-related effects (see next section).

\begin{figure*}[t]
    \centering
    \includegraphics[width=0.9\linewidth]{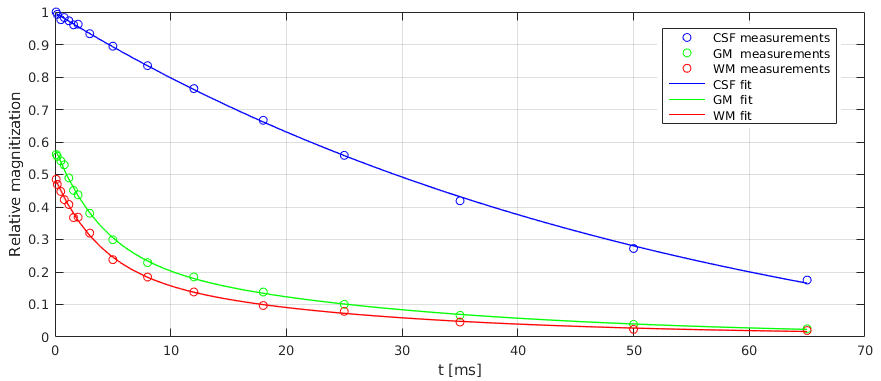}
    \caption{Bi-exponential T$^*_2$ relaxation fitting from the multi-TE phantom data. See section \ref{sec:methods_phantom} for details.}
    \label{fig:T2}
    \vspace{-12pt}
\end{figure*}

\subsubsection{Trajectory \textit{PSF}}

\textbf{Figure \ref{fig:trajectory_total}.a-b} shows the results from PSF simulations for a 2D radial trajectory. PSFs have the same amplitude at any resolution, and FWHM$\sim$1.4 mm only changes by 0.01 mm from one resolution to another, demonstrating that a 2D radial trajectory introduces only a some small amount of blurring regardless of the chosen resolution.  

\subsubsection{Combined blurring function \textit{B}}

\textbf{Figure \ref{fig:bi_total}.c-d} and \textbf{Figure \ref{fig:trajectory_total}.c-d} show the combined effect of decay-related blurring function $f$ and radial trajectory PSF. In the case of variable readout duration (\textbf{Figure \ref{fig:bi_total}.c} and \textbf{Figure \ref{fig:trajectory_total}.c}), the radial trajectory had a significant impact on blurring only at short readout durations (up to 24\% increase in FWHM), whereas its effect was wiped out at longer readout durations (less than 1\% change in FWHM compared to decay-related blurring functions only). In the case of variable dwell time (\textbf{Figure \ref{fig:bi_total}.d} and \textbf{Figure \ref{fig:trajectory_total}.d}), an insignificant difference in FWHM is observable between resolutions (less than 2\% and 3\% for the shortest and longest dwell times, respectively).

\vfill

\subsection{Phantom experiments}

\subsubsection{Phantom relaxation times}

The results of T$^*_2$ fitting (see section \ref{sec:methods_phantom}) are shown in \textbf{Figure \ref{fig:T2}}. The transverse relaxation in CSF followed a mono-exponential decay with $T^*_{2,short}$ = $T^*_{2,long}$ = 54.5 ms. The T$^*_2$ relaxation in the GM and WM compartments followed a bi-exponential decay. For the GM compartment, $T^*_{2,short}=$ 4 ms and $T^*_{2,long}=$ 26.4 ms with $a_{short}=$ 55.7\% and $a_{long}=$ 44.3\%, while for the WM compartment, $T^*_{2,short}=3.9$ ms and $T^*_{2,long}=22.1$ ms with $a_{short}=$ 57.4\% and $a_{long}=$ 42.6\%. These values align with values reported in the literature \cite{madelin2013, boada1994}.

\subsubsection{First experiment: Variable readout duration}

\begin{figure*}[t]
    \centering
    \includegraphics[width=0.9\textwidth]{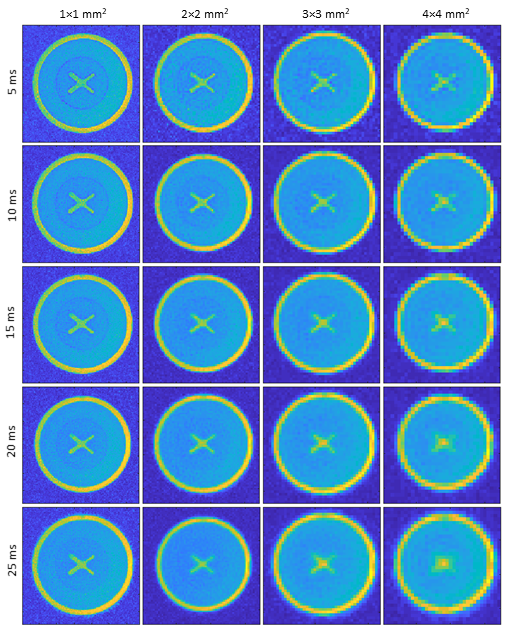}
    \caption{Phantom sodium MRI scans performed at different in-plane resolutions (columns) and readout durations (rows).}
    \label{fig:readout}
\end{figure*}

\textbf{Figure \ref{fig:readout}} shows the results of the first phantom experiment and displays the images with columns corresponding to 4 in-plane resolutions and rows to different readout durations from 5 to 25 ms. For the highest resolution (1$\times$1 mm\textsuperscript{2}), no significant signs of blurring were visible when using a 5 ms readout duration: the image appeared sharp and the walls between the compartments were clearly visible. At longer readout durations, the blurring effect slowly started to manifest itself, as can be noticed in the interface of CSF/GM and CSF/WM compartments. At 20 ms and 25 ms readout durations, these regions became blurrier, as well as the edge between the outer CSF compartment and the air. The border between the GM and WM compartments started to fade as well, and the compartment wall was no longer perceivable. Similar transformations took place at 2$\times$2 mm\textsuperscript{2} resolution but happened "faster" (at shorter readout durations). 

At lower resolutions of 3$\times$3 and 4$\times$4 mm\textsuperscript{2}, a visible blurring effect was already present when using a 5 ms readout duration, e.g., the border between GM and WM compartments was barely visible. When using a 25 ms readout, the blurring deformed the shape of the phantom's central compartment imitating ventricles. The blurring effect was stronger at lower resolutions because the signals from two different compartments were mixed. In contrast, border at the outer surface, interfacing with air, remained more clear. In addition to blurring, low resolution images also exhibited Gibbs artifacts that degraded image quality even further. 

Overall, these images demonstrated that at lower resolutions, the blurring effect was exacerbated and became more prevalent with the increase of readout duration in accordance with the simulation results.

\subsubsection{Second experiment: Short dwell time}

\begin{figure*}[t]
    \centering
    \includegraphics[width=1.0\textwidth]{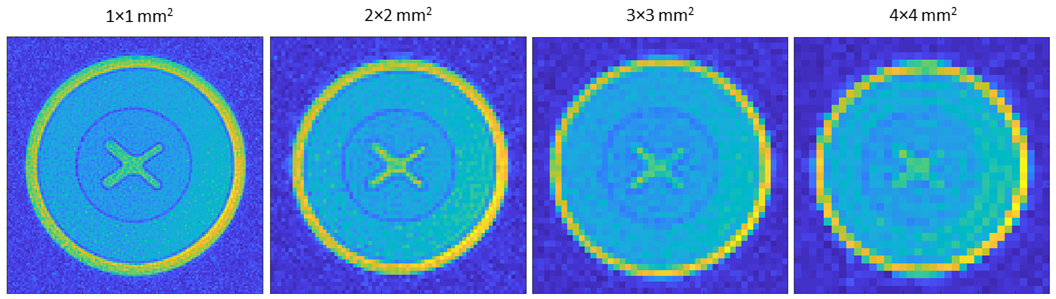}
    \caption{Phantom sodium MRI scans performed at different in-plane resolutions and a same short dwell time.}
    \label{fig:dt}
    \vspace{-12pt}
\end{figure*}

Results of second experiment are displayed in \textbf{Figure \ref{fig:dt}}. Even at very short readout durations ($<$ 1 ms), where the effect of fast \ttwo decay was much less prominent, the images look very different at different resolutions. As expected, the highest-resolution images showed sharp edges and clean curves, while low-resolution images appeared pixelated with fuzzy edges between the phantom compartments, and strong Gibbs artifacts. 

\subsection{In vivo experiment}

\begin{figure*}[b]
    \centering
    \includegraphics[width=0.9\textwidth]{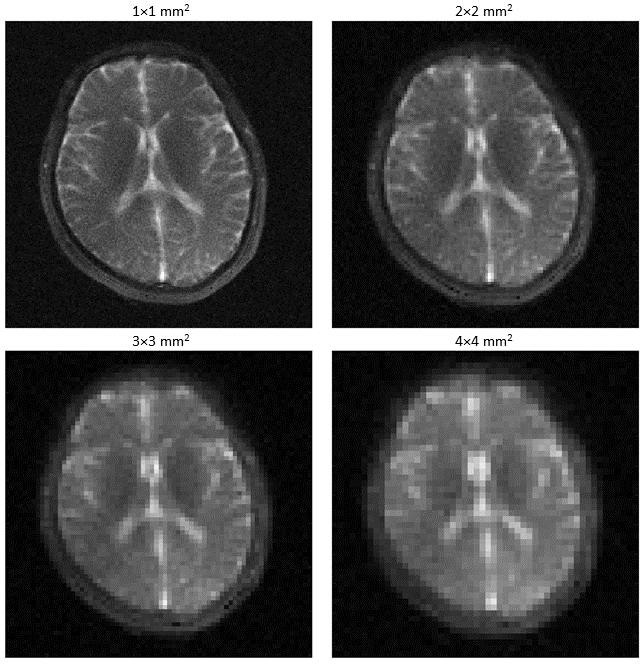}
    \caption{In vivo sodium MRI scans performed with a 15 ms readout duration at four different resolutions.}
    \label{fig:invivo}
    \vspace{-12pt}
\end{figure*}

\textbf{Figure \ref{fig:invivo}} shows the in vivo results. The scan with the highest resolution yielded a sharp brain image with a great amount of detail. Although no clear border between WM and GM could be observed, the CSF in the ventricles and outer parts of the brain was well defined. As expected, with the reduction of resolution (increased pixel size), the images became much blurrier, and the border between WM/GM and CSF faded away at lower resolutions. The decay in all tissues and fluids lead to strong blurring. In particular, the intensified CSF blurring resulted in increased brightness in WM and GM.

\section{Discussion}

The most important observation is that short \ttwo blurring at longer readout durations is strongly exacerbated by low resolution (low k\textsubscript{max}), while it is much less severe at long readout durations but high resolution (high k\textsubscript{max}). As expected, the combined blurring function became wider with the increase of readout duration, however the rate of its widening was smaller for higher resolutions.

Low resolution images exhibit other types of blurring independent of fast transverse decay. The most plausible explanation is that this blurring was the result of partial volume effect, when one voxel contains nuclei with different spin dynamics, originating from different tissues. For example, tissue fractions with long \ttwo will dominate later in the readout, thus weighting the details in those voxels towards the long \ttwo components, even if these only constitute a tiny fraction of the voxel. 

Gradient non-linearity and eddy current effects can also induce blurring in non-Cartesian acquisitions \cite{tao2015}. In each of our experiments, the same radial trajectory was used. Therefore, the only difference was the amplitude of the readout gradient. Ramping up to larger gradient amplitudes could lead to minor differences. Nevertheless, these effects are expected to be smaller for radial than for spiral like trajectories that continuously modulate the gradient amplitudes. 

Therefore, the hypothesis that visible blurring in sodium images is mainly caused by fast \ttwo decay and long readouts appears to be incomplete. It is true in case of high-resolution imaging and long readout acquisitions. However, in practical settings, when most sodium scans are performed at low resolution, the mix of two effects leads to an even stronger blurring effect. Therefore, the choice of resolution should be given a higher importance when aiming to resolve structural details in sodium images.  

Besides increased blurring effect from fast signal decay, low resolution has two other problems: Gibbs ringing and partial volume effects. As our phantom has sharp boundaries between compartments, the Gibbs rings stand out much more than in vivo. Although the brain has more subtile contours and interlacement of GM and WM, Gibbs ringing could still contribute to obfuscate tissue boundaries. Some solutions were proposed to reduce this Gibbs ringing in diffusion MRI \cite{veraart2016}, as well sodium MRI \cite{stobbe2008}. Yet, most commonly, simple post-acquisition Hanning or Hamming filters are used, adding even more blurring. Furthermore, eliminating partial volume effects is still an open-ended question. Both can be avoided by scanning at high resolution, but at the expense of longer acquisition times (more averages) to compensate for the loss of SNR.


Indeed, although our experiments show that high resolution scans yield sharper images with reduced \ttwo blurring and less artifacts, low SNR becomes a vexed problem. A larger number of spokes (or averages) can provide higher SNR but lead to impractical scan times in vivo. Many groups prefer to perform a true 3D acquisition that intrinsically provides a slightly better SNR. However, there are potentially more effective solutions. For example, compressed sensing \cite{lustig2007,madelin2012} can be used as a denoising method leveraging non-sparsity of the noise. Simultaneous acquisition of multinuclear \hhna MR data \cite{yu2020simultaneous,yu2022simultaneous} in combination with joined reconstruction techniques, such as those used in PET/MRI \cite{judenhofer2008}, could also be a path towards high resolution sodium images. Other promising solutions are machine learning based denoising \cite{zhang2017,koppers2019} and super-resolution \cite{yang2019,rodriguez2023super}.

Previously, a major part of \na MRI research was ultimately focused on shortening TE \cite{boada1997a,jereric2004, gai2015,konstandin2015,nielles2007,rahmer2006} to increase SNR and/or shortening readout durations \cite{boada1997b, konstandin2011} to reduce \ttwo decay-induced blurring without compromising SNR. While this led to considerable improvements, blurring remains a challenge. We hope that our study helps to better understand the nature of short \ttwo blurring in sodium images and encourage efforts to increase resolution despite SNR challenges. 

\section{Conclusion}

In conclusion, in this paper, we refined the common hypothesis that blurring in sodium MRI is mainly caused by \ttwo decay. In particular, we demonstrated that low resolutions strongly exacerbate the effect of \ttwo blurring. Therefore, despite its low SNR penalty, aiming for higher resolution sodium images is still expected to help resolve more detailed structures and allow more accurate measurements on quantitative data by strongly reducing partial volume effects due to short \ttwo blurring.  

\clearpage

\onecolumn

\newpage

\newpage

\twocolumn
\bibliography{biblio}

\begin{thebibliography}{10}
\expandafter\ifx\csname url\endcsname\relax
  \def\url#1{\texttt{#1}}\fi
\expandafter\ifx\csname urlprefix\endcsname\relax\def\urlprefix{URL }\fi
\providecommand{\bibinfo}[2]{#2}
\providecommand{\eprint}[2][]{\url{#2}}

\bibitem{madelin2013}
\bibinfo{author}{Madelin, G.} \& \bibinfo{author}{Regatte, R.}
\newblock \bibinfo{title}{{Biomedical applications of sodium MRI in vivo}}.
\newblock \emph{\bibinfo{journal}{Journal of Magnetic Resonance Imaging}} \textbf{\bibinfo{volume}{38}}, \bibinfo{pages}{511--529} (\bibinfo{year}{2013}).

\bibitem{boada1994}
\bibinfo{author}{Boada, F.}, \bibinfo{author}{Christensen, J.}, \bibinfo{author}{Huang‐Hellinger, F.}, \bibinfo{author}{Reese, T.} \& \bibinfo{author}{Thulborn, K.}
\newblock \bibinfo{title}{{Quantitative in vivo tissue sodium concentration maps: the effects of biexponential relaxation}}.
\newblock \emph{\bibinfo{journal}{Magnetic Resonance in Medicine}} \textbf{\bibinfo{volume}{32}}, \bibinfo{pages}{219--223} (\bibinfo{year}{1994}).

\bibitem{fessler2003}
\bibinfo{author}{Fessler, J.} \& \bibinfo{author}{Sutton, B.}
\newblock \bibinfo{title}{{Nonuniform fast Fourier transforms using min-max interpolation}}.
\newblock \emph{\bibinfo{journal}{IEEE Transactions on Signal Processing}} \textbf{\bibinfo{volume}{51}}, \bibinfo{pages}{560--574} (\bibinfo{year}{2003}).

\bibitem{jereric2004}
\bibinfo{author}{Jerecic, R.} \emph{et~al.}
\newblock \bibinfo{title}{{ECG-gated \textsuperscript{23}Na-MRI of the human heart using a 3D-radial projection technique with ultra-short echo times}}.
\newblock \emph{\bibinfo{journal}{Magnetic Resonance Materials in Physics, Biology and Medicine}} \textbf{\bibinfo{volume}{16}}, \bibinfo{pages}{297--302} (\bibinfo{year}{2004}).

\bibitem{boada1997a}
\bibinfo{author}{Boada, F.}, \bibinfo{author}{Gillen, J.}, \bibinfo{author}{Shen, G.}, \bibinfo{author}{Chang, S.} \& \bibinfo{author}{Thulborn, K.}
\newblock \bibinfo{title}{{Fast three dimensional sodium imaging}}.
\newblock \emph{\bibinfo{journal}{Magnetic Resonance in Medicine}} \textbf{\bibinfo{volume}{37}}, \bibinfo{pages}{706--715} (\bibinfo{year}{1997}).

\bibitem{staroswiecki2010}
\bibinfo{author}{Staroswiecki, E.} \emph{et~al.}
\newblock \bibinfo{title}{{In vivo sodium imaging of human patellar cartilage with a 3D cones sequence at 3 T and 7 T}}.
\newblock \emph{\bibinfo{journal}{Journal of Magnetic Resonance Imaging}} \textbf{\bibinfo{volume}{32}}, \bibinfo{pages}{446--451} (\bibinfo{year}{2010}).

\bibitem{gai2015}
\bibinfo{author}{Gai, N.}, \bibinfo{author}{Rochitte, C.}, \bibinfo{author}{Nacif, M.} \& \bibinfo{author}{Bluemke, D.}
\newblock \bibinfo{title}{{Optimized three‐dimensional sodium imaging of the human heart on a clinical 3T scanner}}.
\newblock \emph{\bibinfo{journal}{Magnetic Resonance in Medicine}} \textbf{\bibinfo{volume}{73}}, \bibinfo{pages}{623--632} (\bibinfo{year}{2015}).

\bibitem{pipe2011}
\bibinfo{author}{Pipe, J.} \emph{et~al.}
\newblock \bibinfo{title}{{A new design and rationale for 3D orthogonally oversampled k‐space trajectories}}.
\newblock \emph{\bibinfo{journal}{Magnetic Resonance in Medicine}} \textbf{\bibinfo{volume}{66}}, \bibinfo{pages}{1303--1311} (\bibinfo{year}{2011}).

\bibitem{konstandin2015}
\bibinfo{author}{Konstandin, S.}, \bibinfo{author}{Krämer, P.}, \bibinfo{author}{Günther, M.} \& \bibinfo{author}{Schad, L.}
\newblock \bibinfo{title}{{Sodium magnetic resonance imaging using ultra-short echo time sequences with anisotropic resolution and uniform k-space sampling}}.
\newblock \emph{\bibinfo{journal}{Magnetic Resonance Imaging}} \textbf{\bibinfo{volume}{33}}, \bibinfo{pages}{319--327} (\bibinfo{year}{2015}).

\bibitem{robson2003}
\bibinfo{author}{Robson, M.}, \bibinfo{author}{Gatehouse, P.}, \bibinfo{author}{Bydder, M.} \& \bibinfo{author}{Bydder, G.}
\newblock \bibinfo{title}{{Magnetic resonance: an introduction to ultrashort TE (UTE) imaging}}.
\newblock \emph{\bibinfo{journal}{Journal of computer assisted tomography}} \textbf{\bibinfo{volume}{27}}, \bibinfo{pages}{825--846} (\bibinfo{year}{2003}).

\bibitem{qin2012}
\bibinfo{author}{Qin, Q.}
\newblock \bibinfo{title}{{Point spread functions of the T2 decay in k-space trajectories with long echo train}}.
\newblock \emph{\bibinfo{journal}{Magnetic Resonance in Medicine}} \textbf{\bibinfo{volume}{30}}, \bibinfo{pages}{1134--1142} (\bibinfo{year}{2012}).

\bibitem{constable1992}
\bibinfo{author}{Constable, R.} \& \bibinfo{author}{Gore, J.}
\newblock \bibinfo{title}{{The loss of small objects in variable TE imaging: implications for FSE, RARE, and EPI}}.
\newblock \emph{\bibinfo{journal}{Magnetic Resonance in Medicine}} \textbf{\bibinfo{volume}{28}}, \bibinfo{pages}{9--24} (\bibinfo{year}{1992}).

\bibitem{zhang2009}
\bibinfo{author}{Zhang, L.} \emph{et~al.}
\newblock \bibinfo{title}{{HASTE sequence with parallel acquisition and T2 decay compensation: application to carotid artery imaging}}.
\newblock \emph{\bibinfo{journal}{Magnetic Resonance Imaging}} \textbf{\bibinfo{volume}{27}}, \bibinfo{pages}{13--22} (\bibinfo{year}{2009}).

\bibitem{deichmann2000}
\bibinfo{author}{Deichmann, R.}, \bibinfo{author}{Good, C.}, \bibinfo{author}{Josephs, O.}, \bibinfo{author}{Ashburner, J.} \& \bibinfo{author}{Turner, R.}
\newblock \bibinfo{title}{{The loss of small objects in variable TE imaging: implications for FSE, RARE, and EPI}}.
\newblock \emph{\bibinfo{journal}{NeuroImage}} \textbf{\bibinfo{volume}{12}}, \bibinfo{pages}{112--127} (\bibinfo{year}{2000}).

\bibitem{oshio1989}
\bibinfo{author}{Oshio, K.} \& \bibinfo{author}{Singh, M.}
\newblock \bibinfo{title}{{A computer simulation of T2 decay effects in echo planar imaging}}.
\newblock \emph{\bibinfo{journal}{Magnetic Resonance in Medicine}} \textbf{\bibinfo{volume}{11}}, \bibinfo{pages}{389--397} (\bibinfo{year}{1989}).

\bibitem{ahn1986}
\bibinfo{author}{Ahn, C.}, \bibinfo{author}{Kim, J.} \& \bibinfo{author}{Cho, Z.}
\newblock \bibinfo{title}{{High-speed spiral-scan echo planar NMR imaging-I}}.
\newblock \emph{\bibinfo{journal}{IEEE Transactions on Medical Imaging}} \textbf{\bibinfo{volume}{5}}, \bibinfo{pages}{2--7} (\bibinfo{year}{1986}).

\bibitem{rahmer2006}
\bibinfo{author}{Rahmer, J.}, \bibinfo{author}{Börnert, P.}, \bibinfo{author}{Groen, J.} \& \bibinfo{author}{Bos, C.}
\newblock \bibinfo{title}{{Three‐dimensional radial ultrashort echo‐time imaging with T2 adapted sampling}}.
\newblock \emph{\bibinfo{journal}{Magnetic Resonance in Medicine}} \textbf{\bibinfo{volume}{55}}, \bibinfo{pages}{1075--1082} (\bibinfo{year}{2006}).

\bibitem{nielles2007}
\bibinfo{author}{Nielles‐Vallespin, S.} \emph{et~al.}
\newblock \bibinfo{title}{{3D radial projection technique with ultrashort echo times for sodium MRI: Clinical applications in human brain and skeletal muscle}}.
\newblock \emph{\bibinfo{journal}{Magnetic Resonance in Medicine}} \textbf{\bibinfo{volume}{57}}, \bibinfo{pages}{74--81} (\bibinfo{year}{2007}).

\bibitem{boada1997b}
\bibinfo{author}{Boada, F.}, \bibinfo{author}{Shen, G.}, \bibinfo{author}{Chang, S.} \& \bibinfo{author}{Thulborn, K.}
\newblock \bibinfo{title}{{Spectrally weighted twisted projection imaging: reducing T2 signal attenuation effects in fast three‐dimensional sodium imaging}}.
\newblock \emph{\bibinfo{journal}{Magnetic Resonance in Medicine}} \textbf{\bibinfo{volume}{38}}, \bibinfo{pages}{1022--1028} (\bibinfo{year}{1997}).

\bibitem{wang2009}
\bibinfo{author}{Wang, L.} \emph{et~al.}
\newblock \bibinfo{title}{{Rapid isotropic 3D‐sodium MRI of the knee joint in vivo at 7T}}.
\newblock \emph{\bibinfo{journal}{Journal of Magnetic Resonance Imaging}} \textbf{\bibinfo{volume}{30}}, \bibinfo{pages}{606--614} (\bibinfo{year}{2009}).

\bibitem{wheaton2004}
\bibinfo{author}{Wheaton, A.} \emph{et~al.}
\newblock \bibinfo{title}{{Sodium magnetic resonance imaging of proteoglycan depletion in an in vivo model of osteoarthritis}}.
\newblock \emph{\bibinfo{journal}{Academic Radiology}} \textbf{\bibinfo{volume}{11}}, \bibinfo{pages}{21--28} (\bibinfo{year}{2004}).

\bibitem{bottomley2016}
\bibinfo{author}{Bottomley, P.}
\newblock \bibinfo{title}{{Sodium MRI in human heart: a review}}.
\newblock \emph{\bibinfo{journal}{NMR in Biomedicine}} \textbf{\bibinfo{volume}{29}}, \bibinfo{pages}{187--196} (\bibinfo{year}{2016}).

\bibitem{stobbe2018}
\bibinfo{author}{Stobbe, R.} \& \bibinfo{author}{Beaulieu, C.}
\newblock \bibinfo{title}{{Calculating potential error in sodium MRI with respect to the analysis of small objects}}.
\newblock \emph{\bibinfo{journal}{Magnetic Resonance in Medicine}} \textbf{\bibinfo{volume}{79}}, \bibinfo{pages}{2968--2977} (\bibinfo{year}{2018}).

\bibitem{nagel2009}
\bibinfo{author}{Nagel, A.} \emph{et~al.}
\newblock \bibinfo{title}{{Sodium MRI using a density‐adapted 3D radial acquisition technique}}.
\newblock \emph{\bibinfo{journal}{Magnetic Resonance in Medicine}} \textbf{\bibinfo{volume}{62}}, \bibinfo{pages}{1565--1573} (\bibinfo{year}{2009}).

\bibitem{romanzetti2014}
\bibinfo{author}{Romanzetti, S.} \emph{et~al.}
\newblock \bibinfo{title}{{Mapping tissue sodium concentration in the human brain: a comparison of MR sequences at 9.4 Tesla}}.
\newblock \emph{\bibinfo{journal}{NeuroImage}} \textbf{\bibinfo{volume}{96}}, \bibinfo{pages}{44--53} (\bibinfo{year}{2014}).

\bibitem{wang2021radially}
\bibinfo{author}{Wang, B.} \emph{et~al.}
\newblock \bibinfo{title}{{A radially interleaved sodium and proton coil array for brain MRI at 7 T}}.
\newblock \emph{\bibinfo{journal}{NMR in Biomedicine}} \textbf{\bibinfo{volume}{34}}, \bibinfo{pages}{e4608} (\bibinfo{year}{2021}).

\bibitem{madelin2014}
\bibinfo{author}{Madelin, G.}, \bibinfo{author}{Xia, D.}, \bibinfo{author}{Walvick, R.}, \bibinfo{author}{Lee, J.} \& \bibinfo{author}{Regatte, R.}
\newblock \bibinfo{title}{{FLORET sodium MRI of articular cartilage in the knee joint at 7T}}.
\newblock In \emph{\bibinfo{booktitle}{Proceedings of International Society for Magnetic Resonance in Medicine}}, vol.~\bibinfo{volume}{22}, \bibinfo{pages}{3743} (\bibinfo{year}{2014}).

\bibitem{noll1992}
\bibinfo{author}{Noll, D.}
\newblock \emph{\bibinfo{title}{{Reconstruction techniques for magnetic resonance imaging}}}.
\newblock Ph.D. thesis, \bibinfo{school}{Stanford University} (\bibinfo{year}{1992}).

\bibitem{tao2015}
\bibinfo{author}{Tao, S.} \emph{et~al.}
\newblock \bibinfo{title}{{NonCartesian MR image reconstruction with integrated gradient nonlinearity correction}}.
\newblock \emph{\bibinfo{journal}{Medical Physics}} \textbf{\bibinfo{volume}{42}}, \bibinfo{pages}{7190--7201} (\bibinfo{year}{2015}).

\bibitem{veraart2016}
\bibinfo{author}{Veraart, J.}, \bibinfo{author}{Fieremans, E.}, \bibinfo{author}{Jelescu, I.}, \bibinfo{author}{Knoll, F.} \& \bibinfo{author}{Novikov, D.}
\newblock \bibinfo{title}{{Gibbs ringing in diffusion MRI}}.
\newblock \emph{\bibinfo{journal}{Magnetic Resonance in Medicine}} \textbf{\bibinfo{volume}{76}}, \bibinfo{pages}{301--314} (\bibinfo{year}{2016}).

\bibitem{stobbe2008}
\bibinfo{author}{Stobbe, R.} \& \bibinfo{author}{Beaulieu, C.}
\newblock \bibinfo{title}{{Advantage of sampling density weighted apodization over postacquisition filtering apodization for sodium MRI of the human brain}}.
\newblock \emph{\bibinfo{journal}{Magnetic Resonance in Medicine}} \textbf{\bibinfo{volume}{60}}, \bibinfo{pages}{981--986} (\bibinfo{year}{2008}).

\bibitem{lustig2007}
\bibinfo{author}{Lustig, M.}, \bibinfo{author}{Donoho, D.} \& \bibinfo{author}{Pauly, J.}
\newblock \bibinfo{title}{{Sparse MRI: The application of compressed sensing for rapid MR imaging}}.
\newblock \emph{\bibinfo{journal}{Magnetic Resonance in Medicine}} \textbf{\bibinfo{volume}{58}}, \bibinfo{pages}{1182--1195} (\bibinfo{year}{2007}).

\bibitem{madelin2012}
\bibinfo{author}{Madelin, G.}, \bibinfo{author}{Chang, G.}, \bibinfo{author}{Otazo, R.}, \bibinfo{author}{Jerschow, A.} \& \bibinfo{author}{Regatte, R.}
\newblock \bibinfo{title}{{Compressed sensing sodium MRI of cartilage at 7T: preliminary study}}.
\newblock \emph{\bibinfo{journal}{Journal of Magnetic Resonance}} \textbf{\bibinfo{volume}{214}}, \bibinfo{pages}{360--365} (\bibinfo{year}{2012}).

\bibitem{yu2020simultaneous}
\bibinfo{author}{Yu, Z.}, \bibinfo{author}{Madelin, G.}, \bibinfo{author}{Sodickson, D.~K.} \& \bibinfo{author}{Cloos, M.~A.}
\newblock \bibinfo{title}{{Simultaneous proton magnetic resonance fingerprinting and sodium MRI}}.
\newblock \emph{\bibinfo{journal}{Magnetic Resonance in Medicine}} \textbf{\bibinfo{volume}{83}}, \bibinfo{pages}{2232--2242} (\bibinfo{year}{2020}).

\bibitem{yu2022simultaneous}
\bibinfo{author}{Yu, Z.} \emph{et~al.}
\newblock \bibinfo{title}{{Simultaneous 3D acquisition of \textsuperscript{1}H MRF and \textsuperscript{23}Na MRI}}.
\newblock \emph{\bibinfo{journal}{Magnetic Resonance in Medicine}} \textbf{\bibinfo{volume}{87}}, \bibinfo{pages}{2299--2312} (\bibinfo{year}{2022}).

\bibitem{judenhofer2008}
\bibinfo{author}{Judenhofer, M.~S.} \emph{et~al.}
\newblock \bibinfo{title}{{Simultaneous PET-MRI: a new approach for functional and morphological imaging}}.
\newblock \emph{\bibinfo{journal}{Nature medicine}} \textbf{\bibinfo{volume}{14}}, \bibinfo{pages}{459--465} (\bibinfo{year}{2008}).

\bibitem{zhang2017}
\bibinfo{author}{Zhang, K.}, \bibinfo{author}{Zuo, W.}, \bibinfo{author}{Chen, Y.}, \bibinfo{author}{Meng, D.} \& \bibinfo{author}{Zhang, L.}
\newblock \bibinfo{title}{{Beyond a gaussian denoiser: Residual learning of deep cnn for image denoising}}.
\newblock \emph{\bibinfo{journal}{IEEE Transactions on Image Processing}} \textbf{\bibinfo{volume}{26}}, \bibinfo{pages}{3142--3155} (\bibinfo{year}{2017}).

\bibitem{koppers2019}
\bibinfo{author}{Koppers, S.}, \bibinfo{author}{Coussoux, E.}, \bibinfo{author}{Romanzetti, S.}, \bibinfo{author}{Reetz, K.} \& \bibinfo{author}{Merhof, D.}
\newblock \bibinfo{title}{{Sodium image denoising based on a convolutional denoising autoencoder}}.
\newblock In \emph{\bibinfo{booktitle}{Bildverarbeitung für die Medizin}}, \bibinfo{pages}{98--103} (\bibinfo{publisher}{Springer Vieweg, Wiesbaden}, \bibinfo{year}{2019}).

\bibitem{yang2019}
\bibinfo{author}{Yang, W.} \emph{et~al.}
\newblock \bibinfo{title}{{Deep learning for single image super-resolution: A brief review}}.
\newblock \emph{\bibinfo{journal}{IEEE Transactions on Multimedia}} \textbf{\bibinfo{volume}{21}}, \bibinfo{pages}{3106--3121} (\bibinfo{year}{2019}).

\bibitem{rodriguez2023super}
\bibinfo{author}{Rodriguez, G.~G.} \emph{et~al.}
\newblock \bibinfo{title}{{Super-resolution of sodium images from simultaneous \textsuperscript{1}H MRF and \textsuperscript{23}Na MRI acquisition}}.
\newblock \emph{\bibinfo{journal}{NMR in Biomedicine}} \textbf{\bibinfo{volume}{36}}, \bibinfo{pages}{e4959} (\bibinfo{year}{2023}).

\bibitem{konstandin2011}
\bibinfo{author}{Konstandin, S.}, \bibinfo{author}{Nagel, A.}, \bibinfo{author}{Heiler, P.} \& \bibinfo{author}{Schad, L.}
\newblock \bibinfo{title}{{Two-dimensional radial acquisition technique with density adaption in sodium MRI}}.
\newblock \emph{\bibinfo{journal}{Magnetic Resonance in Medicine}} \textbf{\bibinfo{volume}{65}}, \bibinfo{pages}{1090--1096} (\bibinfo{year}{2011}).

\end{thebibliography}

\section*{Funding}

The research reported in this article was supported by the NIH/NIBIB grant R01EB026456, and performed under the rubric of the Center for Advanced Imaging Innovation and Research, a NIBIB Biomedical Technology Resource Center (P41EB017183).

\end{document}